
\font\titolo=cmbx10 scaled\magstep0

\magnification=\magstep1
\font\myit=cmti7 scaled\magstep2

\parskip 3truemm
\hsize=140 truemm
\vsize=210 truemm
\hoffset=15 truemm

\def\mn{\medskip \noindent}
\def\bi{\bigskip \indent}
\def\bn{\bigskip \noindent}

\def\nostrocostrutto#1\over#2{\mathrel{\mathop{\kern 0pt \rlap
  {\raise.2ex\hbox{$#1$}}}
  \lower.9ex\hbox{\kern-.190em $#2$}}}
\def\lsim{\nostrocostrutto < \over \sim}

\def\aleq{\buildrel < \over \sim}

\def\noi1{noi1}

\def\sv{< \sigma_{ann} v >}

\def\bi{$\tilde B$}
\def\wi{$\tilde W_3$}
\def\c{\chi}
\def\den{\Omega_{\chi} h^2}
\def\m{m_{\chi}}

\nopagenumbers

\bn

\rightline{DFTT 37/93}
\rightline{August 1993}
\bn
\bn
\centerline{\titolo On the Neutralino as Dark Matter Candidate.}
\centerline{\titolo I. Relic Abundance.}
\bn
\mn
\centerline{A. Bottino, V. de Alfaro, N. Fornengo, G. Mignola, M. Pignone}
\bn
\mn
\centerline{\myit {Dipartimento di Fisica Teorica
dell'Universit\`a di Torino}}
\centerline{\myit {and INFN, Sezione di Torino, Italy}}
\centerline{\myit {Via P. Giuria 1, 10125 Torino, Italy}}
\bn
\bn
\bn
\centerline{\bf Abstract}
\medskip
\noindent
The neutralino relic abundance is evaluated for a wide range of the
neutralino mass,
${\rm 20\  GeV} \leq m_\chi \leq {\rm 1\  TeV}$, by taking into account the
full set of final states in the neutralino-neutralino annihilation.
The analysis is performed in the Minimal SuSy Standard Model; it is
not restricted by stringent GUT assumptions but only constrained by
present experimental bounds.  We also discuss phenomenological aspects
which are employed in the companion paper (II. Direct Detection) where
the chances for a successful search for dark matter neutralino are
investigated.

\vfill
\eject

\footline{\hfill}
\headline{\hfil-- \folio\ --\hfil}
\pageno=1

{\bf 1. Introduction.}

   The general features of cosmological structures, as they are observed
and understood at present, lead to the conclusion that large amount of
the matter in our Universe is in the form of Cold Dark Matter (CDM).
This circumstance has recently prompted new detailed investigations
about the neutralino ($\chi$), since this SuSy particle appears to be the
most favorite candidate for CDM.
   Here we analyse one of its basic properties, the relic abundance, by
extending a previous investigation of ours [1]
 (which was confined to
neutralino masses below the W-boson mass) to a much wider
range of neutralino masses:\break
20 GeV $\leq m_\c \leq$ 1 TeV for the
most general neutralino composition. Furthermore we take into account
the whole set of exchange diagrams and final states in the $\c-\c$
annihilation process which is the fundamental ingredient in the
evaluation of the relic abundance.
In addition, radiative corrections to the Higgs boson masses
as well as to the
relevant coupling constants are appropriately included in our
evaluations.

In the present investigation the theoretical framework is
represented by the Minimal Supersymmetric Standard Model (MSSM);
only one standard GUT assumption is employed with the purpose of simplifying
the phenomenological discussion.

   The analysis presented here differs from the previous ones [2-8] in at
least one of the features mentioned above. In particular the most recent
analyses on this subject by other authors  are mainly based upon GUT
schemes which also include supergravity; this automatically implies
model-dependent relationships between the various masses which come into
play. Here we prefer to consider a more flexible scheme where unknown
masses are not {\it a priori} fixed, but are only constrained by present
experimental bounds.

   We also wish to emphasize that we do not restrain our attention
to regions of the parameter space where the neutralino would provide
by itself the total amount of required CDM, but we rather widely
explore the parameter space with the aim of investigating the
chances to detect
the neutralino as a dark matter candidate by direct or indirect
searches. In fact regions where the detection event rates are higher do
not necessarily coincide with the locations in parameter space where the
relic density is larger, since stronger coupling of the neutralino with
matter may compensate for partial depletion in the neutralino local
density. Indeed actual detection of dark matter neutralino
would be an achievement of paramount interest even if the neutralino
does not exhaust our need for CDM.
   For these reasons, whereas in the present paper (hereafter called I)
we discuss the neutralino relic abundance, in the companion paper (II)
which follows this one, we analyse the problem of its direct detection.
The way of presenting our results in paper I is mainly shaped according
to the needs for the applications discussed in paper II.

\bigskip
{\bf 2. Minimal SuSy Standard Model.}

   Our theoretical framework is the MSSM, with the standard definition
of neutralino as the lowest-mass linear combination of photino, zino and
higgsinos,

$$\eqalign{\chi=a_1\tilde\gamma + a_2\tilde Z+a_3\tilde H_1^0
+a_4\tilde H_2^0}\eqno(2.1)$$
where $\tilde \gamma$ and $\tilde Z$ are linear combinations
of the U(1) and SU(2) neutral gauginos, \bi \ and \wi,

$$\eqalign{\tilde \gamma &=\cos\theta_W\tilde B+\sin\theta_W\tilde W_3,\cr
\tilde Z&=-\sin\theta_W\tilde B+\cos\theta_W\tilde W_3,\cr}\eqno(2.2)$$

\noindent
$\theta_W$ being the Weinberg angle. As usual, $\c$ is assumed to be the
lightest supersymmetric particle (LSP) and then stable if R--parity is
conserved. Its mass as well as its composition depend
on the parameters: $M_1$, $M_2$ (masses of {\bi}  and of \wi, respectively),
$\mu$ (Higgs mixing parameter) and $\tan \beta = v_u/v_d$ ($v_u$ and $v_d$
being the v.e.v.'s which give masses to up-type and down-type quarks).
   A standard procedure in current literature is to embed the
MSSM in GUT, so that a relationship between $M_1$ and $M_2$ follows:
$M_1=5/3 \tan^2 \theta_W \simeq 0.5~M_2$. Relaxing this assumption
may modify some neutralino properties
in a significant way; this has been discussed by some authors in the
restricted case $\m<m_W$ [9,10].

In this paper we will report our results under
the assumption that the GUT-induced relation between $M_1$ and $M_2$
holds, as this is a most natural hypothesis.
Extensions of the present analysis due to the relaxation of this
relationship will be presented in a forthcoming paper [11].

   As mentioned above, in the present analysis also high values of $\m$
are considered (up to 1 TeV). Thus, in order to deal with a parameter
space large enough to contain various compositions for $\c$ at any value
of $\m$, we have taken wide ranges for $M_2$ and
$\mu$ : 20 GeV $\leq M_2 \leq$ 6 TeV,
20 GeV $\leq |\mu| \leq$ 3 TeV. The scatter plots which are presented in
section 3 are obtained by varying $M_2$ and $\mu$ in these ranges.
As far as $\tan \beta$ is concerned,
in order to cover a wide range for it we will
representatively choose the following values:
$\tan \beta = 2$, $\tan \beta = 8$
(or rather the range $ 6 \leq \tan \beta \leq 10$
in a number of extended scatter plots) and $\tan\beta = 20$.
It is worth noticing that the ranges chosen for the parameters
automatically disregard regions of the parameters space which have been
excluded by LEP.

   To illustrate the general features of the neutralino we give in
Fig.1 a representation of its mass and composition in the $M_2-\mu$ plane
at $\tan \beta = 8$. This figure clearly shows that the neutralino
composition tends to be very pure (either pure higgsino or pure gaugino,
depending on whether $M_2 > 2 |\mu|$ or $M_2< 2 |\mu|$)
as $M_2$ and $\mu$ increase. This property, already discussed in
previous works, is due to the fact that in the
neutralino $4 \times 4$ mass matrix the higgsino sector asymptotically
decouples from the gaugino sector as $M_2$, $|\mu| \gg M_Z$.
Before this asymptotic regime turns on, and so for $M_2$, $2 |\mu|$ $\lsim$
a few hundreds GeV (neutralino masses $\lsim$ 100 GeV),  mixed
higgsino--gaugino configurations are important. This property plays a
crucial role in some of the conclusions we will draw about the
neutralino relic abundance and the detection event rates.

   Effective interaction properties of the neutralino depend sensitively
on the three physical neutral Higgs bosons of the MSSM: the two
CP-even bosons: h,H (of masses $m_h$, $m_H$ with $m_H>m_h$) and the CP-odd
one: A (of mass $m_A$). Radiatively corrected relationships between these
masses are used here. Thus, taking $m_h$ as an independent parameter, $m_A$
and $m_H$ are functions of $m_h$, $\tan \beta$, $m_t$ (mass of the top quark)
and
$\tilde m$ (mass of the top scalar partners, taken as degenerate). To represent
our results we will take two values for $m_h$: $m_h$ = 50 GeV (which entails
$m_A \sim 50$ GeV) and $m_h = 80$ GeV (this implies $m_A = 83$ GeV for
$\tan \beta = 8$ and $m_A = 170$ GeV for $\tan \beta = 2$). \
The top quark mass has
been taken at the value $m_t = 150$ GeV.
The sfermion masses will be discussed in the next section.

\bigskip
{\bf 3. Relic Abundance.}

   The neutralino relic abundance $\den$ ($\Omega$ is the density
parameter of the Universe: $\Omega=\rho/\rho_c$,
$\rho_c = 1.88\times 10^{-29}h^2~{\rm g}~{\rm cm}^{-3},~ 0.4\leq h\leq 1$)
has been evaluated employing the usual formula [12]

$$ \Omega_\chi h^2 =
2.13 \times 10^{-11} \left ({T_{\c} \over T_\gamma} \right)^3
\left (T_\gamma \over {2.7 K} \right )^3 N_F^{1/2} \left ({\rm GeV}^{-2}
\over {a x_f + {1 \over 2} b x_f^2} \right ). \eqno(3.1) $$

\noindent
Here $T_\gamma$ is the present temperature of the microwave background,
$T_\c/T_\gamma$ is the reheating factor of the photon temperature as
compared to the neutralino temperature, $x_f = T_f/\m \simeq 1/20$
 where $T_f$ is the neutralino freeze--out temperature.
The $x_f$-dependent
expression in the denominator of Eq.(3.1) represents the
integration from $T_f$ down to the
present temperature of the thermally averaged quantity
$< \sigma_{ann} v> = a+bx$
where $\sigma_{ann}$ is the $\c-\c$ annihilation cross-section, $v$ the
relative velocity and $x=T/\m$.
   In the evaluation of the annihilation cross-section we have
considered the whole set of final states: 1) fermion--antifermion pair,
2) pair of neutral Higgs bosons, 3) pair of charged Higgs bosons,
4) one Higgs boson-one gauge boson, 5) pair of gauge bosons
($W^+ W^-$, $Z Z$). For the final state 1), the following
diagrams have been considered: Higgs--exchange diagrams and Z--exchange
diagram in the s--channel, $\tilde f$--exchange diagrams in the t--channel.
For the
final states 2--5) we have taken Higgs-exchange and Z--exchange diagrams
in the s--channel, and either neutralinos (the full set of the four mass
eigenstates)
or chargino exchange in the t--channel, depending on the electric charges
of the final particles.
    As for the sfermion masses, we have considered two extreme cases:
one with mass values as low as possible, compatibly with the present
experimental bound and with the assumption that $\c$ is the LSP, and a
second one where all sfermions are very massive. The lower limit that
we have conservatively used for the sfermion masses is the LEP bound of
{}~45 GeV, since the CDF limit [13] on the squark masses does not appear
to be consistent with a massive neutralino [14].

Thus we have considered the two cases:
1) $m_{\tilde f} = 1.2~ m_\c$,  when $m_\c > 45$ GeV; $m_{\tilde f} = 45$ GeV
otherwise, except for the mass
of the top scalar partner (the only one relevant to radiative corrections)
which has been taken $\tilde m = 3$ TeV ; 2) $m_{\tilde f} = 3$ TeV
for all sfermions.

    Let us now turn to the presentation of our results. In Fig.s 2-3 we
report, in the $M_2 - \mu$ plane, the dominance of various final states
 at $\tan \beta = 8$, $m_h = {\rm 50\ GeV}$ and for the two representative
values for $m_{\tilde f}$. Dominance of a particular final state
means that this
channel weighs for at least a factor of 5 over the other states
in the quantity $\sv$.
Fig. 2, which refers to small values of $m_{\tilde f}$, shows that in large
regions of the $M_2$, $\mu$ plane the $f \bar f$ and the two gauge bosons final
states dominate. However it has to be stressed that for $\mu>0$ in a
significant
region of the parameter space, where mixed gaugino--higgsino configurations
occur, there is dominance of one Higgs boson--one gauge boson final state.
This feature was absent in Ref. [3], where this contribution was
estimated to be subdominant. The one Higgs boson--one gauge boson final states
are even more important when all the $m_{\tilde f}$ masses are set at 3 TeV
(see Fig.3); in fact, a drastic increase in $m_{\tilde f}$ has the obvious
consequence of a suppression of
the $\tilde f$--exchange amplitude, and this significantly depresses the
$f \bar f$ final states. On the contrary, dominance of the two Higgs
boson final states is limited to a few isolated points in the parameter
space, as is illustrated in Fig.s 2--3. These general features remain almost
unaltered if we move from  $\tan \beta = 8$ to smaller values,
$\tan \beta \sim 2$.

    Let us now discuss the neutralino relic abundance. In Fig. 4 we
report our results in the form of scatter plots obtained by varying $M_2$,
$\mu$ in the parameter space previously defined and by varying
$\tan \beta$ in the range $6 \leq \tan \beta \leq 10$. $\den$
versus $\m$ is shown for three
different types of neutralino compositions: higgsino dominance
(dominance here means 90\% or more), gaugino
dominance, maximal higgsino--gaugino mixing
(i.e., $0.45 \le a_1^2+a_2^2 \le 0.55$).
In Fig. 4a, (higgsino dominance) some characteristic features
are very clearly displayed. A number of pronounced dips (and of sharp
falls off) in $\den$ reflect the presence of poles (and the opening
of new thresholds) in the annihilation cross section. In sequence we
have: at $\m \sim 25$ GeV the $h$ and the $A$ poles,
at $\m \sim 45$ GeV the $Z$
pole, at $\m \sim 90$ GeV the threshold for the $\c-\c$ annihilation into
channels $W^+ W^-$ and $ZZ$. In the case of gaugino dominance
(see Fig.4b) sfermion--exchange amplitudes provide large
contributions to the annihilation cross section, with the effect of
depressing the neutralino relic abundance as compared to the case of
higgsino dominance
for $m_\chi \aleq {\rm 90\ GeV}$. At higher $m_\chi$ values
the gaugino--dominated compositions give a larger $\den$, since
here the $f \bar f$ final state is dominant, but somewhat hampered by the
running values of the $\tilde f$ mass: $m_{\tilde f} = 1.2~ m_\c$.
Compositions with large mixings (see Fig.4c-d) entail rather
low values of $\den$, due to the substantial contribution to the annihilation
provided by Higgs--exchange and $\tilde f$--exchange.
Nevertheless, these neutralino
configurations contribute significantly to the event rates for direct
neutralino search (as discussed in paper II).
   The previous discussion should make the features of Fig.5 quite
transparent. In fact here high values for $m_{\tilde f}$ force the neutralino
relic density of the gaugino--dominated configurations to be large, by
inhibiting the $\tilde f$--exchange amplitude. This same
mechanism is the reason for the depletion in the $f \bar f$ final state
dominance that we notice in the plot of Fig.3 when we compare it with
the one in Fig.2.

   Fig.s 6--7 show the neutralino relic abundance when $\tan \beta$ is small:
   ($\tan \beta = 2$, for definiteness) and $m_h = 80$ GeV.
Higgsino--dominated configurations at small $\m$ (below the thresholds
for $W^+ W^-$ and $ZZ$) display large values of
$\den$, since now,
smaller values of $\tan \beta$ and larger values of
Higgs boson masses, both have
the effect of suppressing the annihilation channels with Higgs
exchanges. As for the gaugino--dominated compositions it is worth
noticing that in the case of large $m_{\tilde f}$ (see Fig.7b)
$\den$ displays a behaviour which is rather common in the context of
some supergravity inspired models. In fact in these schemes it frequently
occurs that theoretical and
phenomenological constraints restrict $\tan \beta$ to very small values
and sfermion masses to high values, with the consequence that the
cosmological requirement $\den < 1$ can only be met at the
Higgs--poles or at the Z--pole. Consequently, particular care has to be
taken in the evaluation of the relic abundance [15] in these models.
In our kind of analysis, fine--tuning of $\m$ with the masses of the
Higgs bosons or of the gauge bosons is not required and would then
appear rather accidental.

A final scatter plot for $\den$ is shown in Fig.8 for $\tan \beta = 20$
and $m_h = {\rm 50 \  GeV}$.

   A word of warning is required about the effects due to the possible
occurrence of an approximate degeneracy (within $ \sim 15 \%$) between the
neutralino and some other SuSy particle. When this happens the
neutralino decoupling mechanism is
enhanced due to the annihilation process involving the neutralino with
the other SuSy particle which is close to it in mass (this process is usually
denoted as coannihilation in the literature) [15,16]. Effects on $\den$
due to coannihilation may be large (one order of magnitude or more,
depending on the nature of the coannihilating particle and on other details
of the theoretical scheme). Apart from the
peculiar and accidental case when for instance a sfermion and the $\chi$
would almost have the same mass, a natural case of approximate
degeneracy occurs in the neutralino-chargino sector.
However, this happens in regions of the parameter space of higgsino
dominance. In paper II it is shown that chances of detecting dark matter
neutralinos rely essentially on mixed or gaugino compositions for
neutralinos. Thus coannihilation does not significantly affect the evaluations
of the event rates presented in paper II.

In conclusion our analysis confirms that the neutralino, even at mass
values higher than the W-mass, may satisfy the attributes required
for a good candidate for CDM. It is also clear that the theoretical
evaluations for neutralino relic abundance only suffer from the lack of
information about some of the particles that naturally come into play,
such as the Higgs bosons and any SuSy object. Only new experimental
inputs can help theory in sharpening its predictions for neutralino
dark matter.

\bigskip
\centerline{*\ \ \  *\ \ \  *}
\bigskip

This work was supported in part by Research Funds of the Ministero
dell'Universit\`a e della Ricerca Scientifica e Tecnologica.

\vfill
\eject
\centerline{\bf References}

\item{[1]}
A.Bottino, V.de Alfaro, N.Fornengo, G.Mignola and S.Scopel,
{\it Astroparticle Phys.} {\bf 1}(1992)61.

\item{[2]} References where $\m$ beyond the W--mass has
been considered are
given in [3-8]. For other papers see, for instance, the references quoted
in [1].
\item{[3]} K.Griest, M.Kamionkowski and M.S.Turner, Phys. Rev. D41(1990)3565
\item{[4]} J.McDonald, K.A.Olive and M.Srednicki, Phys. lett. B283(1992)80.
\item{[5]} P.Gondolo, M.Olechowski and S.Pokorski, MPI-Ph/92-81 preprint.
\item{[6]} M.Drees and M.M.Nojiri, Phys. Rev. D47(1993)376.
\item{[7]} P.Nath and R.Arnowitt, CTP-TAMU-66/92 preprint;\hfill\break
J.L.Lopez, D.V.Nanopoulos, and K.Yuan, CTP-TAMU-14/93 preprint.
\item{[8]} R.G.Roberts and L.Roszkowski, Phys. Lett. B309(1993)329.
\item{[9]} K.Griest and L.Roszkowski, Phys. Rev. D46(1992)3309.
\item{[10]} S.Mizuta, D.Ng and M.Yamaguchi, Phys. Lett. B300(1993)96.
\item{[11]} A.Bottino, V.de Alfaro, N.Fornengo, G.Mignola and M.Pignone,
to appear.
\item{[12]} J.Ellis, J.S.Hagelin, D.V.Nanopoulos, K.Olive and M.Srednicki,
Nucl. Phys. B238(1984)453.
\item{[13]} F.Abe et al. (CDF Collaboration), Phys. Rev. Lett. 69(1992)3439.
\item{[14]} see also H.Baer, X.Tata and J.Woodside, Phys. Rev. D44(1991)207
for a discussion on cascade decays whose inclusion weakens
the bounds of Ref.[13].
\item{[15]} K.Griest and D.Seckel, Phys. Rev. D43(1991)3191.
\item{[16]} S.Mizuta and M.Yamaguchi, Phys. Lett. B298(1993)120.

\vfill
\eject

\centerline{\bf Figure Captions}

{\bf Figure 1}.
Isomass curves and composition lines for neutralino in the
$M_2-\mu$ plane for $\tan \beta = 8$.
Dashed lines are
lines of constant $\chi$ mass ($m_\chi$ = 30 \ GeV, 100 \ GeV,
300 \ GeV and 1 \ TeV). Solid lines refer to constant gaugino
fraction $f_g$ in the neutralino composition
($f_g = a_1^2 +a_2^2$): $f_g$ = 0.99, 0.9,
0.5, 0.1 and 0.01.

\bigskip

{\bf Figure 2.}
Final states dominance regions in $<\sigma_{ann} v>$,
for $\tan \beta =8$ and $m_h = {\rm 50 \ GeV}$.
Sfermion masses are given by:
$m_{\tilde{f}}$ = 45 \ GeV, when $m_\chi <$ 45 \ GeV;
$m_{\tilde{f}} = 1.2 \ m_\chi$ otherwise (except for the SuSy partners
of the top quark whose common mass is set at $\tilde{m}$ = 3 \ TeV).
Dominance region for a particular channel is defined as the region where
that channel dominates over the other ones by a factor of five at least.
Different regions are marked as follows:
heavy dots for $f \bar{f}$ final state,
horizontal lines for gauge boson pair final state,
diamonds for mixed Higgs boson - gauge boson final state,
squares for Higgs boson pairs final state.
In regions marked with light dots, no dominance of a particular
channel occurs.

\bigskip

{\bf Figure 3.}
Same as in Figure 2, with all sfermion masses fixed at
$m_{\tilde{f}} = {\rm 3 \ TeV}$.

\bigskip

{\bf Figure 4.}
Scatter plots for neutralino relic abundance $\Omega_\chi h^2$
as a function of the neutralino mass $m_\chi$.
$M_2$ and $\mu$ are varied in the ranges
${\rm 20 \ GeV} \leq M_2 \leq {\rm 6 \ TeV}$ and
${\rm 20 \ GeV} \leq |\mu| \leq {\rm 3 \ TeV}$;
$\tan\beta$ is varied in the range $6 \le \tan\beta \le 10$;
lightest scalar Higgs boson mass is $m_h = {\rm 50 \ GeV}$;
sfermion masses are taken as in Figure 2.
(a) and (b) refer to neutralino compositions which
are dominantly higgsino ($a_1^2+a_2^2 \le$ 0.1) or
dominantly gaugino ($a_1^2+a_2^2 \ge$ 0.9), respectively;
(c) refers to the case when higgsino and
gaugino components are maximally mixed
($0.45 \le a_1^2+a_2^2 \le 0.55$), for positive $\mu$;
(d) the same as in (c), for negative $\mu$.

\bigskip

{\bf Figure 5.}
Same as in Figure 4(a and b), except for sfermion masses fixed at
$m_{\tilde{f}} = {\rm 3 \ TeV}$.

\bigskip

{\bf Figure 6.}
Scatter plots for neutralino relic abundance $\Omega_\chi h^2$
as a function of the neutralino mass $m_\chi$.
$M_2$ and $\mu$ are varied in the ranges
${\rm 20 \ GeV} \leq M_2 \leq {\rm 6 \ TeV}$ and
${\rm 20 \ GeV} \leq |\mu| \leq {\rm 3 \ TeV}$;
$\tan\beta$ is fixed at the value $\tan\beta=2$;
lightest scalar Higgs boson mass is $m_h = {\rm 80 \ GeV}$;
sfermion masses are taken as in Figure 2.
(a) and (b) refer to neutralino compositions which
are dominantly higgsino ($a_1^2+a_2^2 \le$ 0.1) or
dominantly gaugino ($a_1^2+a_2^2 \ge$ 0.9), respectively;
(c) refers to the case when higgsino and
gaugino components are maximally mixed
($0.45 \le a_1^2+a_2^2 \le 0.55$), for positive $\mu$;
(d) the same as in (c), for negative $\mu$.

\bigskip

{\bf Figure 7.}
Same as in Figure 6(a and b), except for sfermion masses fixed at
$m_{\tilde{f}} = {\rm 3 \ TeV}$.

\bigskip

{\bf Figure 8.}
Same as in Figure 6, except for
$\tan\beta= 20$ and $m_h = {\rm 50 \ GeV}$.

\bye